\documentclass{jfm}
\usepackage{graphicx}
\usepackage{amsmath,amssymb}

\DeclareMathOperator{\sgn}{sgn}

\newcommand{\uztilde}{\tilde{u}_z}
\usepackage{xcolor}

\shorttitle{Stratified Resistive Tearing Instability}
\shortauthor{S. J. Hopper, T. S. Wood and P. J. Bushby}

\title{Stratified Resistive Tearing Instability}

\author{Scott J. Hopper\aff{1}
  \corresp{\email{s.hopper@newcastle.ac.uk}},
  Toby S. Wood\aff{1}
 \and Paul J. Bushby\aff{1}}

\affiliation{\aff{1}School of Mathematics, Statistics and Physics, Newcastle University, Newcastle upon Tyne, NE1 7RU, UK}

\begin{document}

\maketitle

\begin{abstract}\label{Abstract}
Resistive tearing instabilities are common in fluids that are highly electrically conductive and carry strong currents. We determine the effect of stable stratification on the tearing instability under the Boussinesq approximation.
Our results generalise previous work that considered only specific parameter regimes,
and we show that the length scale of the fastest growing mode depends non-monotonically on the stratification strength. We confirm our analytical results by solving the linearised equations numerically, and we discuss whether the instability could operate in the solar tachocline.
\end{abstract}


\section{Introduction}\label{sec:Intro}

In magnetised fluids of large but finite conductivity, tearing-type instabilities frequently arise in regions where one component of the magnetic field changes sign. In such regions field lines can easily reconnect, forming structures of closed field lines known as plasmoids or ``magnetic islands" \citep{Ji-etal22}.
Tearing instabilities have long been studied in the context of plasma confinement experiments \citep{Furth63}, the Earth's magnetotail \citep{Coppi66} and the solar atmosphere \citep{SomVern93}, and more recently in diverse areas such as neutron stars \citep{Lyutikov03, Wood-etal14, GourgHoller16}, post-main-sequence stars \citep{Kaufman22}, and generic magneto-hydrodynamic (MHD) turbulence \citep{BoldyrevLoureiro18}.
In each of these contexts, the crucial property of the instability is that it leads to reconnection, releasing energy from the magnetic field,
even in fluids that are close to the ideal MHD limit.

The simplest geometry in which to study tearing instability is the ``sheet pinch'' \citep{Furth63}, in which a thin and flat current sheet is embedded within a large-scale magnetic field.
\citet{Furth63} considered an MHD 
model with asymptotically small resistivity, and showed that the tearing instability can be described analytically by solving the ideal equations in the bulk of the fluid, solving the resistive equations in a boundary layer within the current sheet, and asymptotically matching these solutions.
Their model included variations in density and resistivity, but to simplify the analysis they assumed that such variations are small,
and they made a further approximation that is valid only if the growth rate is slower than resistive diffusion across the boundary layer;
this latter approximation came to be known as the ``constant-$\psi$ approximation''.
Later studies have generalised this analysis to include various additional physical effects,
but the majority of analytical results to date have been dependent on the same constant-$\psi$ approximation.
Under this approximation,
the existence of the instability can be demonstrated,
but it is not possible to obtain the full dispersion relation or to identify the fastest growing mode.
An alternative approach is to solve the full set of (linear or nonlinear) equations numerically \citep[e.g.][]{Dahlburg83,Califano99,Attico00,Landi08,Jelinek17, Kaufman22}. However, with this approach it is not possible to understand how the instability behaves in the asymptotic regime of high conductivity, which is often the regime of most physical interest.

\citet{Coppi76} and \citet{PegSchep86} showed that,
in the simplest case (with constant density and resistivity),
the boundary-layer problem can be solved analytically, which allows the full dispersion relation to be obtained and the fastest growing mode to be identified.
In fact, this boundary-layer problem had already been solved earlier by \citet{GibsonKent71}, and generalised to include density stratification by \citet{BaldwinRoberts72}.
Unfortunately, their work seems to have been largely overlooked in the plasma physics community,
and the full solution of the unstratified case is usually attributed to \citeauthor{Coppi76} \citep[e.g.][]{BoldyrevLoureiro18}.

In early studies, density stratification
was often included because of its analogy with the effect of stellarator curvature \citep[e.g.][]{Furth63,Johnson63}.
In the present work,
our motivation is the solar tachocline, which is a strongly stably stratified layer within the Sun
that is believed to harbour a strong toroidal magnetic field.
There are many instabilities that may play a role in the dynamics of the tachocline,
such as magnetic buoyancy \citep{Parker55, Hughes07, Gilman18},
magneto-rotational instability (MRI) \citep{BalbusHawley91, ParfreyMenou07, Ogilvie07, KaganWheeler14, Gilman18}, clamshell and
tipping-type instabilities \citep{Cally03},
as well as non-magnetic, shear-driven instabilities \citep{SpiegZahn70, Garaud01}.
However, the possibility of tearing instability has seldom been mentioned in this context,
outside of a few studies \citep{JiDaughton11,Lewis22},
none of which considered the effect of stratification.
Recently, however, it has been suggested that the tachocline contains a toroidal field whose sign oscillates in the radial direction,
as a result of inward diffusion of the cyclic dynamo field from the overlying convective envelope \citep{ForgacsDajkaPetrovay01,Barnabe-etal17}. {Whether such a field configuration is compatible with the Sun's interior rotation,
and with the solar dynamo cycle,
remains a matter of debate
\citep[e.g.][]{Gough07,Matilsky-etal22}.
In any case, such a}
field configuration seems likely to be subject to tearing instability,
unless it is suppressed by the tachocline's stabilising stratification.
This motivates us to investigate the degree to which stable stratification affects the instability,
and hence to assess whether it can arise within the solar tachocline.

In this paper we re-derive the boundary-layer solution of \citet{BaldwinRoberts72}
and use it to fully describe the effect of stratification on the tearing instability.
We identify several different parameter regimes depending on the degree of stratification,
and demonstrate that previous results are reproduced in particular asymptotic limits.
We then confirm our results by solving the linearised equations numerically.
The plan of the paper is as follows.
Our mathematical model is defined and the equations are linearised in section~\ref{sec:Setup}.
We derive analytical solutions valid in the bulk and boundary layer in sections~\ref{sec:Bulk} and \ref{sec:Boundary}, respectively.
In section~\ref{sec:AsympMatching} we use asymptotic matching to obtain an implicit dispersion relation,
and describe its properties in the asymptotic limit of large conductivity.
In section~\ref{sec:NumericalWork} we solve the linearised equations numerically,
validating the analytical results and
quantifying the effect of finite conductivity and domain size on the instability. {The results are applied to the solar tachocline in section~\ref{sec:Tachocline_Application},
where we also consider the effect of thermal diffusion.
We summarise our findings in section~\ref{sec:Conclusion}.}

\section{The sheet pinch model}\label{sec:Setup}

We consider an inviscid, stably stratified, Boussinesq fluid under the MHD approximation:
\begin{gather}
 \frac{\partial\boldsymbol{u}}{\partial t} + \boldsymbol{u}\boldsymbol{\cdot}\boldsymbol{\nabla} \boldsymbol{u} = -\frac{1}{\rho_0}\boldsymbol{\nabla}P + \theta\boldsymbol{e}_{z} + \frac{1}{4\pi\rho_{0}}(\boldsymbol{\nabla}\times\boldsymbol{B})\times\boldsymbol{B}, \label{eq:Momentum_Strat}
\\
  \label{eq:MHD_Strat}
    \frac{\partial\boldsymbol{B}}{\partial t} = \boldsymbol{\nabla}\times(\boldsymbol{u} \times\boldsymbol{B}) + \eta\nabla^{2}\boldsymbol{B},
 \\
  \label{eq:Maxwell2_Strat}
    \boldsymbol{\nabla}\boldsymbol{\cdot}\boldsymbol{B} = 0,
  \\
  \label{eq:Incompressibility_Strat}
    \boldsymbol{\nabla}\boldsymbol{\cdot}\boldsymbol{u} = 0, 
   \\
  \label{eq:Buoyancy_Strat}
    \frac{\partial\theta}{\partial t} + \boldsymbol{u} \boldsymbol{\cdot}\boldsymbol{\nabla}\theta = -N^{2}u_{z},
\end{gather}
where $\boldsymbol{e}_{z}$ is the unit vector in the $z$ (vertical) direction, $P$ is the pressure, $\rho_0$ is the (constant) reference density, $\boldsymbol{u}$ is the fluid velocity, with vertical component $u_z$, $\boldsymbol{B}$ is the magnetic field (measured in Gaussian c.g.s.\ units), $\eta$ is the magnetic diffusivity, $N$ is the (constant) buoyancy frequency, and $\theta$ is
the buoyancy variable.

{In these equations we have included magnetic diffusion,
because it is essential for tearing instability to operate,
but we have neglected the diffusion of both momentum and temperature.
As we shall see, this simplification makes it possible to solve the boundary-layer problem analytically.
The effect of momentum diffusion is generally to reduce the growth rate, as has been demonstrated in previous works \citep[e.g.][]{Porcelli87,Tenerani-etal15}.
The effect of thermal diffusion, which is certainly important in the solar tachocline,
will be addressed in section~\ref{subsec:TDiff}.
Unsurprisingly, the main consequence of including thermal diffusion is to reduce the stabilising effect of the stratification.
}

\subsection{Background state}
\label{subsec:Background}
\begin{figure} 
    \centering
    \includegraphics[scale=0.35]{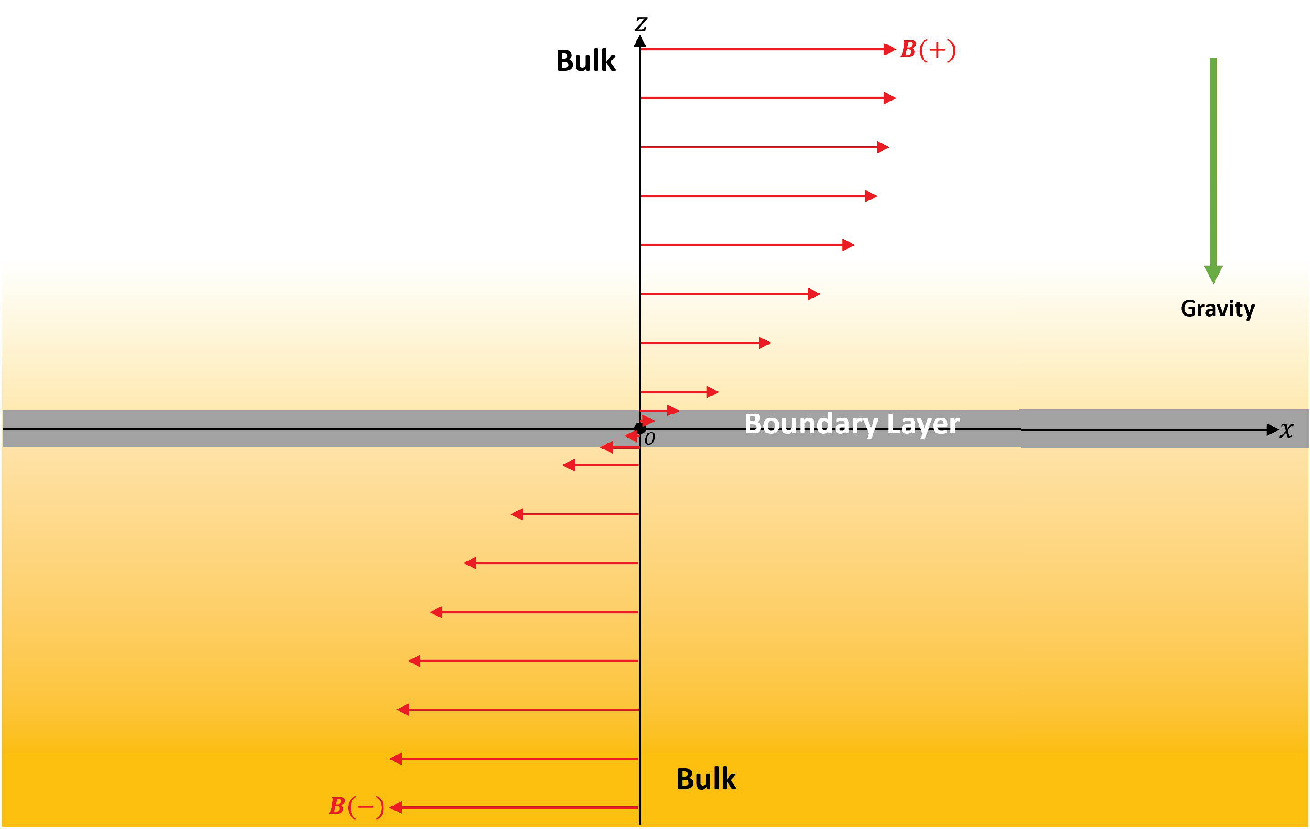}
    \caption{Initial background state configuration. Horizontal arrows indicate the strength and direction of the magnetic field, and the colour gradient indicates the stable density stratification.
    Tearing instability arises from magnetic diffusion within an internal boundary layer, indicated by a thin grey strip around the $x$-axis.}
    \label{fig:Setup}
\end{figure}

As illustrated in figure~\ref{fig:Setup},
we consider a background state at rest with a magnetic field $\boldsymbol{B} = (B(z),0,0)$ in Cartesian coordinates.
The crucial condition for tearing instability to occur is that the sign of $B(z)$ reverses for some value of $z$, which we will take to be $z=0$ without loss of generality.
The fastest-growing tearing modes are generally found to be invariant in the direction of the electric current
{\citep[e.g.][]{Furth63}}, which in our case is the $y$-direction,
and so for simplicity we will only consider two-dimensional perturbations in the $xz$-plane.
Such modes are insensitive to any component of the background field in the $y$-direction, and so we have taken this to be zero without loss of generality.
For simplicity we will assume that $B(z)$ is an odd function, which implies that the
solutions
of the linear perturbation problem have either even or odd symmetry.

In what follows, for definiteness we will generally adopt the so-called Harris field,
\begin{equation}
  B(z) = \sqrt{4\pi\rho_0}\beta\ell \tanh(z/\ell),
  \label{eq:Harris}
\end{equation}
{\citep[named for][]{Harris62}}
where the constant $\beta$ quantifies the Alfv\'enic shear in the current sheet,
which has thickness $\ell$.
We then define the Lundquist number as
\begin{equation}
  S \equiv \frac{\beta\ell^2}{\eta}\,.
  \label{eq:Lundquist}
\end{equation}
However, it is straightforward to generalise our results to other choices for the background field.

We note that the background state is not strictly a steady solution of the induction equation~\eqref{eq:MHD_Strat}
in the presence of finite resistivity.
However, we are concerned with the regime in which $\eta$ is very small, in the sense that $S \gg 1$,
and the slow diffusion of the background state can be neglected provided that the instability grows on a timescale much shorter than the bulk diffusion time, $\ell^2/\eta$.
In what follows, it is often convenient to measure quantities in units defined by the background field,
and in particular to use $\ell$ as the length scale and $1/\beta$ as the time scale,
in which case the condition for self-consistency of our instability analysis is naturally written as $\sigma/\beta \gg 1/S$,
where $\sigma$ is the growth rate.
In terms of these natural units, the strength of the stratification can be expressed as a ``magnetic Richardson number'',
\begin{equation} \label{eq:MagRich}
  R_B \equiv \frac{N^2}{\beta^2},
\end{equation}
named by analogy with the hydrodynamic Richardson number
that determines the stability of shear flows.

\subsection{Linear perturbations} \label{subsec:LinPerts}

As mentioned above, we anticipate that the fastest growing tearing mode will be invariant in the direction of the electric current,
which is the $y$-direction for our choice of background state.
We therefore consider small perturbations to the background state in the form
\begin{equation*}
    a(x,z,t) = \mathrm{e}^{\mathrm{i} kx + \sigma t}\,\hat{a}(z),
\end{equation*}
where $a(x,z,t)$ represents a perturbation to any of the variables,
$k$ is the wavenumber, $\sigma$ is the growth rate,
and $\hat{a}(z)$ is the perturbation eigenfunction.
The linearised versions of equations~\eqref{eq:Momentum_Strat}--\eqref{eq:Buoyancy_Strat} can then be reduced to the following pair of coupled ordinary differential equations
for the $z$-components of $\boldsymbol{u}$ and $\boldsymbol{b}$:
\begin{gather}
    \label{eq:FullODE_I}
     \sigma^2\hat{u}_{z}'' - (\sigma^2 + N^2)k^2\hat{u}_{z}
     = \frac{\mathrm{i} k\sigma}{4\pi\rho_0}\left[B\hat{b}_{z}'' - \left(k^{2} B + B''\right)\hat{b}_{z}\right],
     \\
     \label{eq:FullODE_II}
     \eta\hat{b}_{z}'' - (\eta k^2 + \sigma)\hat{b}_{z} = - \mathrm{i} kB\hat{u}_{z},
\end{gather}
where a prime ($'$) denotes a derivative with respect to $z$.

Following the method introduced by \citet{Furth63},
we will consider separately the bulk and boundary-layer regions of the domain,
solving a leading-order asymptotic approximation to
equations \eqref{eq:FullODE_I} and \eqref{eq:FullODE_II} in each region.
The solutions will then be connected by asymptotic matching, in order to obtain a dispersion relation for the tearing instability.
For simplicity, we will take the bulk domain to be infinite in extent,
with a background field $B(z)$ that is bounded in magnitude.
In that case, the only physically meaningful solutions are those for which the perturbations are also bounded at infinity, and this will serve as our boundary condition.
(Other solutions would correspond to an injection of energy from infinity, rather than an instability arising internally.)
Furthermore, if $B(z)$ is an odd function (as we shall assume throughout)
then we generally expect the fastest growing solution to have an even $\hat{b}_{z}(z)$, and therefore an odd $\hat{u}_{z}(z)$, along with a real growth rate.
{These properties, which are frequently assumed in studies of tearing instability \citep[e.g.][]{Coppi76},
will be confirmed in Appendix~\ref{sec:Frobenius}.}

\section{The bulk solution} \label{sec:Bulk}
Within the bulk of the domain we neglect the diffusion of the field,
and therefore omit the terms in equation~\eqref{eq:FullODE_II} involving $\eta$;
this approximation requires the growth rate, $\sigma$,
{to exceed the rate of diffusion, i.e.~$\sigma \gg \eta(k^2 + 1/\ell^2)$.
Having neglected these terms, it is straightforward to combine Equations~\eqref{eq:FullODE_I} and \eqref{eq:FullODE_II} into a single equation:
\begin{equation}
  \sigma^2\hat{u}_{z}'' - (\sigma^2 + N^2)k^2\hat{u}_{z}
     = \frac{k^2}{4\pi\rho_0}\left[(B^2\hat{u}_{z}')' - k^{2}B^2 \hat{u}_{z}\right].
\end{equation}
We further assume that the growth rate of the instability is small compared to the Alfv\'en frequency, i.e.~$\sigma \ll \beta \ell k$,
which allows us to neglect the terms involving $\sigma^2$ on the left-hand side of this equation.
The validity of all these assumptions
will be verified once the growth rate is known.
Thus, we finally arrive at the bulk equation
\begin{equation} \label{eq:Bulk_ODE}
    \left(B^2\hat{u}_{z}'\right)' = \left(k^2B^2 + 4\pi\rho_0 N^2\right)\hat{u}_{z}.
\end{equation}
}
The same equation was obtained by \citet{Furth63},
although they assumed that the $N^2$ term was small enough to be neglected in the bulk.
When this term is not negligible, which is the regime of interest in the present work,
the nature of the bulk equation changes subtly but significantly
\citep[e.g.][]{Johnson63}. 
We note that $z=0$, which is the location of the boundary layer, is a regular singular point of equation~\eqref{eq:Bulk_ODE}.
This is expected, since it is within the boundary layer that resistivity (which we have neglected in the bulk equation)
is required in order to regularise the solutions. As we approach $z=0$, where we assume that the background field
is of the form $B(z) \simeq \sqrt{4\pi\rho_0}\beta z$,
the general solution of equation~\eqref{eq:Bulk_ODE} is a superposition of two Frobenius power series with the exponents
$- \tfrac{1}{2} \pm \sqrt{\tfrac{1}{4} + N^2/\beta^2} = - \tfrac{1}{2} \pm \sqrt{\tfrac{1}{4} + R_B}$. Therefore unlike the unstratified case, in which $\hat{u}_{z}$ can be written as a series in integer powers of $z$, in the stratified problem $\hat{u}_{z}$ is a superposition of power series with generally incommensurate exponents.
In particular, we can anticipate that a difficulty will arise when the stratification is increased to the critical value 
$R_B = 3/4$, because then the Frobenius exponents will differ by $2$, implying that logarithmic terms must appear in the solution. As we shall see, this means that the usual procedure for matching the bulk solution to the boundary-layer solution fails.
We note that non-integer exponents in the solution can also result from other physical processes, such as geometric curvature \citep{Glasser75} and the Hall effect \citep{Attico00}, but the consequences for the asymptotic matching problem have not been fully recognised in the literature.

The above considerations are generic for any sensible choice of background field $B(z)$. In the particular case of the Harris field \eqref{eq:Harris}, the bulk equation can be solved in closed form by first making the transformation $T = \tanh^2(z/\ell)$ to put it into hypergeometric form:
{
\begin{equation}
  \frac{\mathrm{d}^2\hat{u}_{z}}{\mathrm{d}T^2} + \left(\frac{3/2}{T} - \frac{1}{1-T}\right)\frac{\mathrm{d}\hat{u}_{z}}{\mathrm{d}T}
  =
  \frac{s^2T + (r^2-\tfrac{1}{4})(1-T)}{4T^2(1-T)^2}\hat{u}_{z}\,,
\end{equation}
 where we have introduced dimensionless parameters $r$ and $s$, which are defined as
\begin{equation}
  \label{eq:r_and_s}
  r \equiv \sqrt{\tfrac{1}{4} + R_B}
  \qquad \mbox{and} \qquad
  s \equiv \sqrt{(k\ell)^2 + R_B}.
\end{equation}
}
The solution that has the required boundedness as $|z| \to \infty$
{can be expressed in terms of the hypergeometric function ${_2F_1}$ as}
\begin{equation}
  \hat{u}_{z} = T^{-\tfrac{1}{4}+\tfrac{1}{2}r}(1-T)^{s/2} {_2F_1}(\tfrac{1}{2}r+\tfrac{1}{2}s-\tfrac{1}{4},\tfrac{1}{2}r+\tfrac{1}{2}s+\tfrac{5}{4},s+1,1-T)\sgn(z),
  \label{eq:Bulk_Solution}
\end{equation}
where we have included a factor of $\sgn(z)$ so that $\hat{u}_{z}(z)$ is an odd function, for reasons explained earlier. 
From here on, the strength of the stratification will generally be measured in terms of the parameter $r$ or $R_B$, rather than $N$.
The unstratified case corresponds to $r=1/2$, and the critical stratification mentioned above corresponds to $r=1$.
From equation~\eqref{eq:Bulk_Solution} it can be confirmed that the bulk solution near $z=0$ has logarithmic behaviour when $r=1$ (or when $r$ is any other integer --- see Appendix~\ref{sec:Frobenius}).

\section{The boundary layer solution} \label{sec:Boundary}

The boundary layer is assumed to be thin, in the sense that its thickness
{(which will be precisely defined later)}
is smaller than both $\ell$ and $1/k$ by a factor of $S$ to some positive power. 
In equations \eqref{eq:FullODE_I} and \eqref{eq:FullODE_II}
we will therefore approximate $\nabla^2 \simeq \partial^2/\partial z^2$ and
\begin{equation*}
    B(z) \simeq \sqrt{4\pi\rho_0}\,\beta z.
\end{equation*}
These approximations result in the boundary-layer equations
\begin{gather}
    \label{eq:Boundary_ODE_I}
     \sigma\hat{u}_{z}'' = \frac{\mathrm{i} k\beta z}{\sqrt{4\pi\rho_0}}\hat{b}_{z}'' + \frac{k^2 N^{2}}{\sigma}\hat{u}_{z},
     \\
     \label{eq:Boundary_ODE_II}
     \sigma\hat{b}_{z} = \sqrt{4\pi\rho_0}\mathrm{i} k\beta z\hat{u}_{z} + \eta\hat{b}_{z}''.
\end{gather}
Because these equations involve four derivatives, but only two factors of $z$, they are more easily solved by working in Fourier space.
Defining the Fourier transform of $\hat{u}_{z}$ as
\[
  \uztilde(\zeta) = \int_{-\infty}^{\infty}\mathrm{e}^{\mathrm{i}\zeta z}\hat{u}_{z}(z)\,\mathrm{d}z\,,
\]
equations~\eqref{eq:Boundary_ODE_I}--\eqref{eq:Boundary_ODE_II}
can be transformed and combined into a single equation,
\begin{equation} \label{eq:Fourier_Boundary_ODE}
    \left[\frac{\sigma\zeta^2}{\beta^2} + R_B\frac{k^2}{\sigma}\right]\tilde{u}_z = k^2 \frac{\mathrm{d}}{\mathrm{d}\zeta}\left(\frac{\zeta^2}{\sigma + \eta\zeta^2}\frac{\mathrm{d}\tilde{u}_z}{\mathrm{d}\zeta}\right).
\end{equation}
This is a second-order ordinary differential equation in $\zeta$,
with an essential singularity at $\zeta \to \infty$.
We are interested in the solution that is exponentially small at infinity,
since only this solution has an inverse Fourier transform.
We note that, because the function $\hat{u}_{z}(z)$ is odd, its transform $\uztilde(\zeta)$ must also be odd.
We will therefore solve equation~\eqref{eq:Fourier_Boundary_ODE} in the domain $0<\zeta<\infty$ and impose antisymmetry about $\zeta=0$.

As well as an essential singularity at $\zeta \to \infty$,
equation~\eqref{eq:Fourier_Boundary_ODE} also has three regular singularities at $\zeta = 0$ and at $\zeta = \pm \mathrm{i}\sqrt{\sigma/\eta}$.
On closer inspection, however, the latter two are found to be only ``apparent'' singularities, suggesting that there is a change of variable that reduces this equation to a solvable form \citep{ShanCrast02}.
Indeed, if we define a new independent variable 
$X = \frac{\sqrt{\eta\sigma}}{\beta k}\zeta^{2}$
{then the boundary-layer equation becomes
\begin{equation}
  4X^{1/2}\frac{\mathrm{d}}{\mathrm{d}X}\left(\frac{X^{3/2}}{X+\lambda}\frac{\mathrm{d}\uztilde}{\mathrm{d}X}\right)
  =
  \left(X + \frac{r^2-\tfrac{1}{4}}{\lambda}\right)\uztilde\,,
\end{equation}
where $r$ is defined as in equation~\eqref{eq:r_and_s} and
\begin{equation} \label{eq:lambda_Def}
    \lambda \equiv \frac{\sqrt{\sigma^3/\eta}}{\beta k}
    = S^{1/2}(\sigma/\beta)^{3/2}(k\ell)^{-1}.
\end{equation}
Following the method of \citeauthor{ShanCrast02},
we can express the desired solution}
in terms of the Tricomi function $U$ as
\begin{align}
    \uztilde = \mathrm{e}^{-X/2}&X^{-\frac{1}{4}+\frac{r}{2}}\left[U\left(\frac{(r+\lambda)^{2}-\frac{1}{4}}{4\lambda},1+r,X\right)\right.
    \nonumber \\
     &\left.+ \left(\frac{(\lambda-\tfrac{1}{2})^{2} - r^{2}}{4\lambda}\right) U\left(\frac{(r+\lambda)^{2}-\frac{1}{4}}{4\lambda} + 1,1+r,X\right)\right]\sgn(\zeta).
     \label{eq:BL_Solution}
\end{align}
This is equivalent to the solution obtained by \citet{BaldwinRoberts72},
and in the unstratified limit, $r\to\tfrac{1}{2}$,
it reduces to the solution obtained by \citet{PegSchep86}.
{From this solution, the thickness of the boundary layer can be defined as the region in which the variable $X$ is of order unity,
which corresponds in $z$-space to a thickness of $(\eta\sigma)^{1/4}/(\beta k)^{1/2}$.
Unfortunately, at this stage we do not know the order of magnitude of the growth rate, $\sigma$, and so the boundary layer is of indeterminate thickness.
However, our earlier assumption that it is much smaller than $\ell$ can now be expressed more precisely as $\sigma/\beta \ll S(k\ell)^2$,
which will be verified \textit{a posteriori}.}

We note, in passing, that our boundary-layer equation~\eqref{eq:Fourier_Boundary_ODE} has the same mathematical form as one that arises when analysing tearing instability in the electron-MHD regime \citep[e.g.][]{Attico00,Wood-etal14}.
However, to our knowledge, the fact that it can be solved analytically has not previously been recognised in that context.

For the purposes of asymptotically matching the boundary-layer solution to the bulk solution,
as done in the following section,
we need only determine the behaviour of the boundary-layer solution for ``large'' values of $z$ (in comparison with the scale of the boundary layer itself),
which is dictated by the behaviour of equation~\eqref{eq:BL_Solution} for $X \ll 1$.
Specifically, we will make use of the result {\citep[see][]{Kammler08}}
\begin{equation} \label{eq:FT_IFT_uz}
  \uztilde(\zeta) \sim |\zeta|^{-\alpha}\sgn(\zeta) \quad \mbox{as $\zeta\to0$}
  \qquad \Longleftrightarrow \qquad
  \hat{u}_{z}(z) \sim \frac{|z|^{\alpha-1}\sgn(z)}{2\mathrm{i}\sin(\tfrac{\pi}{2}\alpha)\Gamma(\alpha)} \quad \mbox{as $|z|\to\infty$}
\end{equation}
to determine the behaviour of the boundary-layer solution for large $z$ from its solution~\eqref{eq:BL_Solution} in Fourier space.

\section{Asymptotic matching} \label{sec:AsympMatching}

Now that we are in possession of analytical solutions valid in the bulk and boundary layer regions,
all that is required is to asymptotically match these two solutions,
and thus arrive at a dispersion relation.
Following \citet{Furth63},
in nearly all previous studies of tearing instability
this has been achieved essentially by matching the coefficients of the two leading-order terms in the series representations for the bulk and boundary-layer solutions.
Our bulk solution~\eqref{eq:Bulk_Solution} for $\hat{u}_{z}(z)$ has the form
\begin{equation}
  \hat{u}_{z} = \sum_{n=0}^{\infty}\left[A_{n}|z|^{-\frac{1}{2}-r+2n} + B_{n}|z|^{-\frac{1}{2}+r+2n}\right]\sgn(z),
  \label{eq:Bulk_Frobenius}
\end{equation}
whereas our boundary-layer solution~\eqref{eq:BL_Solution},
after transforming back into $z$-space, has the form
\begin{equation}
  \hat{u}_{z} = \sum_{n=0}^{\infty}\left[a_{n}|z|^{-\frac{1}{2}-r-2n} + b_{n}|z|^{-\frac{1}{2}+r-2n}\right]\sgn(z),
  \label{eq:BL_Frobenius}
\end{equation}
where the coefficients $A_n$, $B_n$, $a_n$ and $b_n$ are known functions of $k$ and $\sigma$.

Matching the first terms in each of the four series, bearing in mind that each solution also allows an arbitrary overall factor, we obtain an implicit dispersion relation:
\begin{equation} \label{eq:CoeffDispRel}
  \frac{A_{0}}{B_{0}}(k) = \frac{a_{0}}{b_{0}}(\sigma,k).
\end{equation}
However, the terms that are matched under this procedure cease to be the leading-order terms when $r \geqslant 1$.
Indeed, if $r=1$ then, as mentioned earlier, both the bulk and the boundary-layer solutions will feature logarithmic terms (of different forms) and hence this matching process cannot be valid in that case. Fortunately, as we shall show,
the tearing instability is strongly suppressed by stable stratification
before this mathematical difficulty arises,
so the standard matching procedure
leading to equation~\eqref{eq:CoeffDispRel}
is adequate for our purposes. 

Taking the values for the coefficients $A_0$, $B_0$, $a_0$ and $b_0$ implied by equations~\eqref{eq:Bulk_Solution} and \eqref{eq:BL_Solution} (see Appendix~\ref{sec:Frobenius}),
we thus obtain the dispersion relation
\begin{align}
  (Sk\ell)^{2r/3} \frac{\Gamma(r)}{\Gamma(-r)} \frac{\Gamma(\tfrac{s}{2}-\tfrac{r}{2}-\tfrac{1}{4})}{\Gamma(\tfrac{s}{2}+\tfrac{r}{2}-\tfrac{1}{4})} \frac{\Gamma(\tfrac{s}{2}-\tfrac{r}{2}+\tfrac{5}{4})}{\Gamma(\tfrac{s}{2}+\tfrac{r}{2}+\tfrac{5}{4})}
  &= \nonumber \\
  \lambda^{r/3} \frac{\Gamma(-r)}{\Gamma(r)}\frac{\lambda+\tfrac{1}{2}+r}{\lambda+\tfrac{1}{2}-r}&\frac{\Gamma\left(\frac{(\lambda+r)^2-\tfrac{1}{4}}{4\lambda}\right)}{\Gamma\left(\frac{(\lambda-r)^2-\tfrac{1}{4}}{4\lambda}\right)}\frac{\sin[\tfrac{\pi}{2}(\tfrac{1}{2}+r)]}{\sin[\tfrac{\pi}{2}(\tfrac{1}{2}-r)]}\frac{\Gamma(\tfrac{1}{2}+r)}{\Gamma(\tfrac{1}{2}-r)}\,.
  \label{eq:DispersionRelation_Full}
\end{align}
Although this result is highly implicit,
we note that the left-hand side of equation \eqref{eq:DispersionRelation_Full}
(which comes from the bulk solution)
is only a function of $k$,
and the right-hand side
(which comes from the boundary-layer solution)
is only a function of $\lambda$,
which itself is related to $\sigma$ and $k$ by equation~\eqref{eq:lambda_Def}.
In fact, for any value of $r \in [\tfrac{1}{2},1)$,
and for any value of $S > 0$, we can prove that
there is always a single fastest growing mode.
First observe that the left-hand side of equation~\eqref{eq:DispersionRelation_Full}
is a positive and monotonically increasing function of $k$ for $0 < k\ell < \sqrt{r + \tfrac{1}{2}}$;
it vanishes as $k\ell \to 0$ and diverges as $k\ell \to \sqrt{r+\tfrac{1}{2}}$.
Meanwhile the right-hand side is a positive and monotonically decreasing function of $\lambda$ for for $0 < \lambda < r + \tfrac{1}{2}$;
it diverges as $\lambda \to 0$ and vanishes as $\lambda \to r+\tfrac{1}{2}$.
Therefore equation~\eqref{eq:DispersionRelation_Full} describes a monotonically decreasing relation between $k$ and $\lambda$ over this range of $k$.
Expressing this result in terms of the growth rate, $\sigma$,
which is related to $\lambda$ by equation~\eqref{eq:lambda_Def},
we find that $\sigma$ vanishes at $k=0$ and at $k=\sqrt{r+\tfrac{1}{2}}/\ell$, and has a unique maximum within this range of $k$,
which corresponds to the fastest growing tearing mode.

In the unstratified limit, $r \to \tfrac{1}{2}$, equation~\eqref{eq:DispersionRelation_Full} reduces to the well-known result
{\citep[e.g.][]{Coppi76,PegSchep86,BoldyrevLoureiro18}}
\begin{equation}
  S^{1/3}\frac{(k\ell)^{4/3}}{1-(k\ell)^2}
  =
  \frac{1-\lambda^2}{\pi\lambda^{5/6}}\frac{\Gamma(\frac{1+\lambda}{4})}{\Gamma(\frac{3+\lambda}{4})}.
  \label{eq:UnstratifiedDispRel}
\end{equation}
In that case, and
in the asymptotic limit $S \to \infty$, the fastest growing mode has $\lambda$ of order unity and $k\ell \ll 1$, such that the left-hand side can be approximated as a power law $\propto k^{4/3}$.
Hence, using the definition of $\lambda$ in equation~\eqref{eq:lambda_Def},
the fastest growing mode has $k\ell \sim S^{-1/4}$ and $\sigma/\beta \sim S^{-1/2}$.

Interestingly, for any non-zero amount of stratification
(i.e.~for any value of $r>1/2$),
the left-hand side of equation~\eqref{eq:DispersionRelation_Full} has a different asymptotic behaviour in the limit $k \to 0$
(and so does the right-hand side in the limit $\lambda \to 0$).
This implies that the instability changes qualitatively in the presence of stratification (at least in the asymptotic limit of large $S$),
and as we show below, several distinct asymptotic regimes arise in the simultaneous limit of $S \to \infty$ and $r \to 1/2$.

\begin{figure}
    \centering
    \includegraphics[scale=0.6]{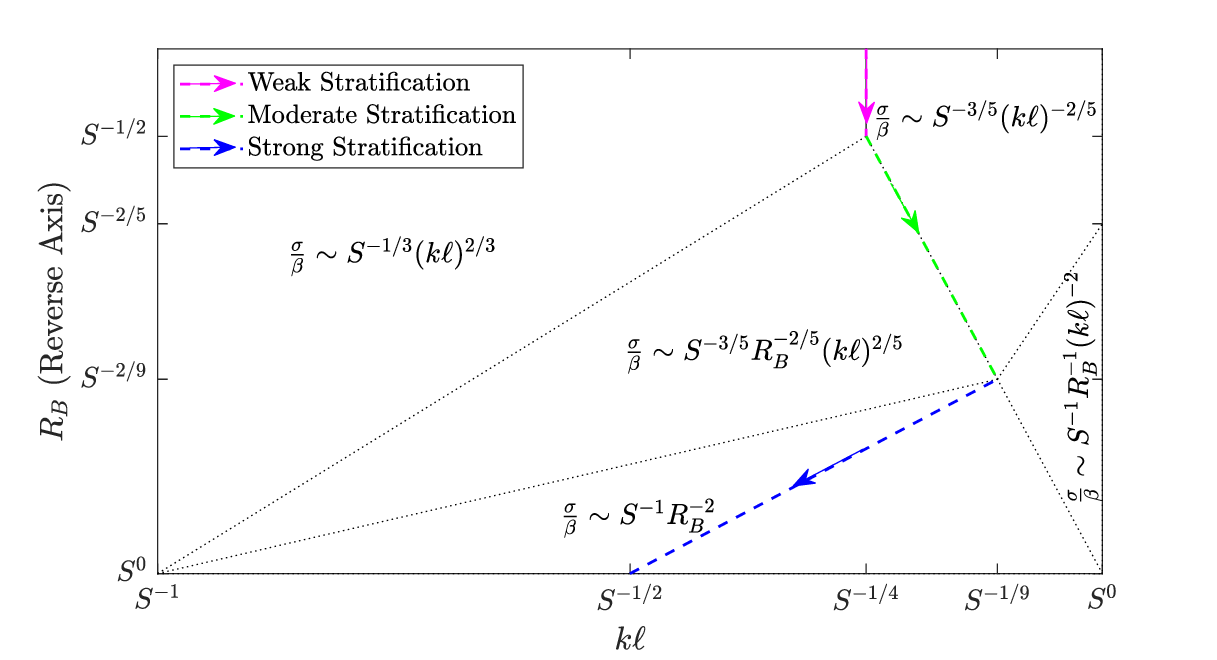}
     \caption{Scaling regimes, {demarcated by dotted lines,} for the growth rate, $\sigma$, in the asymptotic limit $S\to\infty$. Coloured arrows track the behaviour of the fastest growing mode as stratification is increased.
     The axes are logarithmic.}
    \label{fig:Regimes}
\end{figure}

\begin{figure}
    \centering
    \includegraphics[scale=0.6]{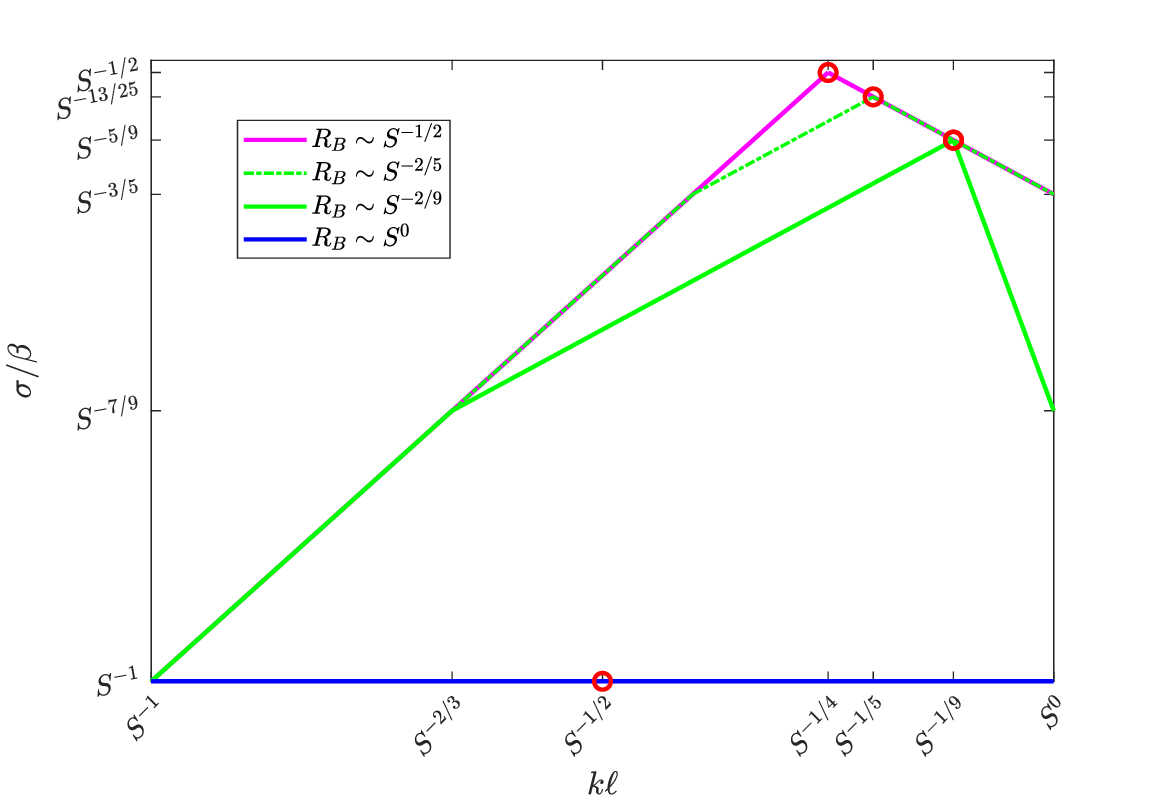}
    \caption{The dispersion relation in the asymptotic limit $S \to \infty$, with $R_B \sim S^{-a}$,
    for $a=1/2$, $2/5$, $2/9$ and $0$.
    The location of the fastest growing mode is indicated with a circle.}
    \label{fig:AsympDispRels}
\end{figure}

Recalling the definition~\eqref{eq:r_and_s} of $r$ in terms of $R_B$,
we note that the limit $r \to 1/2$ is equivalent to $R_B \to 0$.
In the following, we will therefore consider the asymptotic limit $S \to \infty$ with $R_B \sim S^{-a}$ for some positive constant $a$. To understand the behaviour of the dispersion relation~\eqref{eq:DispersionRelation_Full} in this limit,
it is helpful to consider how the left- and right-hand sides behave for different ranges of $k$ and $\lambda$, respectively.
For $R_B \ll 1$, the left-hand side of equation~\eqref{eq:DispersionRelation_Full} is of order
\begin{subequations}
\begin{flalign}
  && S^{1/3}R_B^{1/2}(k\ell)^{1/3} &\qquad \mbox{for $k\ell \ll R_B^{1/2}$}& \\
  && S^{1/3}(k\ell)^{4/3} &\qquad \mbox{for $R_B^{1/2} \ll k\ell \ll 1$},&
\end{flalign}
\end{subequations}
while the right-hand side is of order
\begin{subequations}
\begin{flalign}
  && R_B^{-1/2}\lambda^{-1/3} &\qquad \mbox{for $\lambda \ll R_B$}& \\
  && \lambda^{-5/6} &\qquad \mbox{for $R_B \ll \lambda \ll 1$}& \\
  && 1 - \lambda &\qquad \mbox{for $\lambda \sim 1$}.&
\end{flalign}
\end{subequations}
From this information, and the fact that
$\lambda \equiv S^{1/2}(\sigma/\beta)^{3/2}(k\ell)^{-1}$,
we can identify how the growth rate, $\sigma$, scales with the wavenumber, $k$, in different regions of the parameter space.
The result is shown in figure~\ref{fig:Regimes}, which illustrates how new regimes arise as the strength of the stratification is increased
(or, equivalently, as the value of the exponent $a$ is decreased).
Qualitative changes to the dispersion relation occur for $a=\tfrac{1}{2}$, $\tfrac{2}{5}$, $\tfrac{2}{9}$ and $0$,
and the form of the function $\sigma(k)$ for each of these critical values is indicated in figure~\ref{fig:AsympDispRels}.
We note that, as we would expect in a stably stratified system,
increasing the strength of the stratification acts to decrease the growth rate (at a given value of $k$).

The assumptions made about the growth rate in {sections~\ref{sec:Bulk} and \ref{sec:Boundary}}
can now be checked, with the aid of figure~\ref{fig:Regimes}.
In particular, we have assumed that $S^{-1}(1 + (k\ell)^2) \ll \sigma/\beta \ll k\ell$
{and that $\sigma/\beta \ll S(k\ell)^2$},
and we find that these assumptions hold in all of the regions plotted,
as long as $k\ell \gg S^{-1}$ and $R_B \ll 1$.

We are primarily interested in how the fastest growing mode changes in the presence of stratification,
and as illustrated by the dashed arrows in figure~\ref{fig:Regimes}
there are three distinct parameter regimes to consider,
which we shall refer to as weakly, moderately and strongly stratified.

\subsection{The weakly stratified regime, $a \geqslant 1/2$} \label{subsec:Weak}

If $a > 1/2$, then in the limit $S \to \infty$ the
entire dispersion relation is essentially identical to the unstratified case, $R_B=0$.
The fastest growing mode thus has $k\ell \sim S^{-1/4}$ and $\sigma/\beta \sim S^{-1/2}$,
as described above.

\begin{figure}
    \centering
    \includegraphics[scale=0.6]{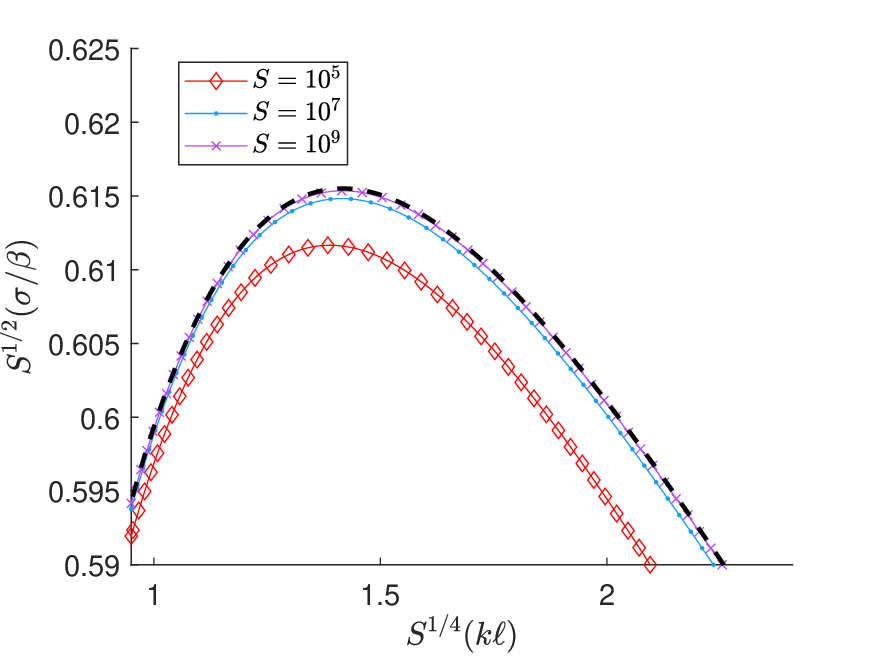}
    \caption{
    The dispersion relation~\eqref{eq:DispersionRelation_Full} for various values of $S$,
    with $R_B = 0.1S^{-1/2}$.
    The dashed curve shows the asymptotic solution~\eqref{eq:a1_2_DispRel} obtained in the limit $S\to\infty$.}
    \label{fig:a=1/2_NoNums}
\end{figure}

In the case $a=1/2$, the effects of stratification begin to affect the fastest growing mode,
and to show how we can introduce a rescaled wavenumber,
$\bar{k} \equiv k\ell/R_B^{1/2}$, which remains of order unity in the limit $S \to \infty$.
In this limit, the dispersion relation~\eqref{eq:DispersionRelation_Full} can be approximated as
\begin{equation}
  (SR_B^2)^{1/3}\,\bar{k}^{1/3}(1 + \bar{k}^2)^{1/2}
  = \frac{1-\lambda^2}{\pi\lambda^{5/6}}\frac{\Gamma(\frac{1+\lambda}{4})}{\Gamma(\frac{3+\lambda}{4})}\,.
  \label{eq:a1_2_DispRel}
\end{equation}
Figure~\ref{fig:a=1/2_NoNums} illustrates how the exact dispersion relation approaches this asymptotic form for increasing values of $S$ in the case where $R_B = 0.1S^{-1/2}$.

\subsection{The moderately stratified regime, $2/9 \leqslant a < 1/2$} \label{subsec:Moderate}

For stronger stratification (i.e.~for smaller values of $a$),
the effect of stratification in the bulk domain suppresses a range of wavenumbers, including the fastest growing mode.
As a result, the fastest growing mode shifts to smaller length scales, with
$k\ell \sim R_B^{1/2}$, and has a reduced growth rate of $\sigma/\beta \sim S^{-3/5}R_B^{-1/5}$.

\begin{figure}
    \centering
    \includegraphics[scale=0.6]{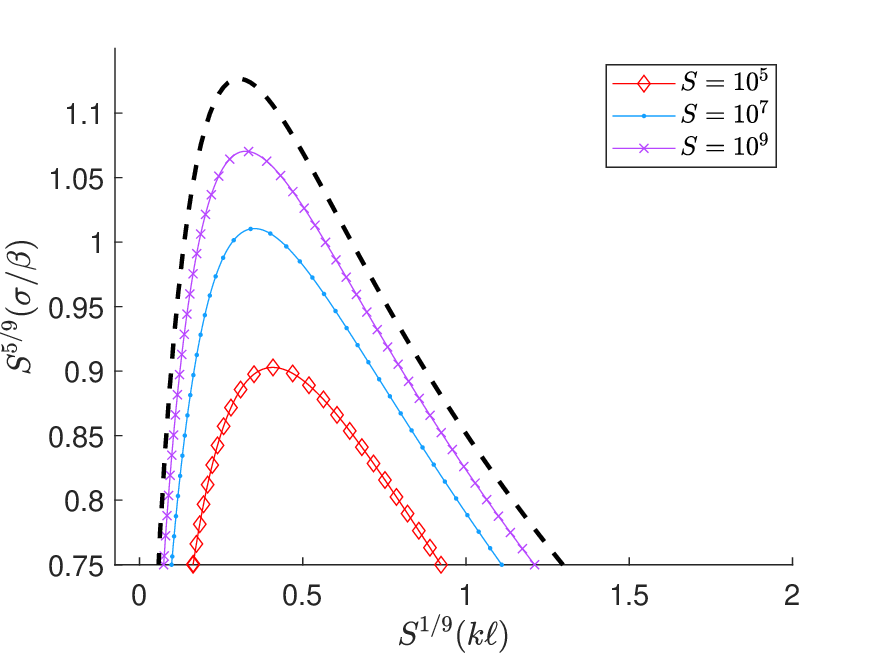}
    \caption{
    The dispersion relation~\eqref{eq:DispersionRelation_Full} for various values of $S$,
    with $R_B = 0.1S^{-2/9}$.
    The dashed curve shows the asymptotic solution~\eqref{eq:a2_9_DispRel} obtained in the limit $S\to\infty$.
    }
    \label{fig:a=2/9_NoNums}
\end{figure}

For $a < 2/5$, the effects of stratification begin to be felt also in the boundary layer, suppressing very small scales,
but this does not affect the fastest growing mode until $a 
\leqslant 2/9$.
In the case $a=2/9$ we can introduce rescaled parameters
$\bar{k} \equiv k\ell/R_B^{1/2}$ and $\bar{\lambda} \equiv \lambda/R_B$ that remain of order unity in the limit $S \to \infty$.
In this limit, the dispersion relation~\eqref{eq:DispersionRelation_Full} can be approximated as
\begin{equation}
  (SR_B^{9/2})^{1/3}\bar{k}^{1/3}(1 + \bar{k}^2)^{1/2}
  = \frac{1}{\pi\bar{\lambda}^{5/6}}\frac{\Gamma(\frac{1+1/\bar{\lambda}}{4})}{\Gamma(\frac{3+1/\bar{\lambda}}{4})}\,.
  \label{eq:a2_9_DispRel}
\end{equation}
Figure~\ref{fig:a=2/9_NoNums} illustrates how the exact dispersion relation approaches this asymptotic form for increasing values of $S$ in the case where $R_B = 0.1S^{-2/9}$.
We note that the right-hand side of equation~\eqref{eq:a2_9_DispRel} is equivalent to the result obtained by \citet{Johnson63} using an extension of the constant-$\psi$ approximation.

\subsection{The strongly stratified regime, $0 \leqslant a < 2/9$} \label{subsec:Strong}

For even stronger stratification the peak in the growth rate broadens, such that we have
$\sigma/\beta \sim S^{-1}R_B^{-2}$
throughout the range $S^{-1}R_B^{-4} \ll k\ell \ll R_B^{1/2}$.
The fastest growing mode,
which can be identified by considering the next-to-leading-order terms in the dispersion relation,
is found at $k\ell \sim S^{-1/2}R_B^{-7/4}$.

In the case $a = 0$, for which $R_B$ is of order unity, the growth rate is very small --- of order $\sigma \sim S^{-1}\beta$.
This is comparable to the rate of diffusion of the magnetic field in the bulk, and so a key assumption of our analysis no longer holds (see section~\ref{sec:Bulk}).
Moreover, as mentioned earlier, as $R_B \to 3/4$ (and so $r \to 1$)
the dispersion relation~\eqref{eq:DispersionRelation_Full} becomes singular (because $\Gamma(-r)$ diverges),
and so our dispersion relation is clearly not valid for $R_B$ of order unity. 

\begin{figure}
    \centering 
    \includegraphics[scale=0.6]{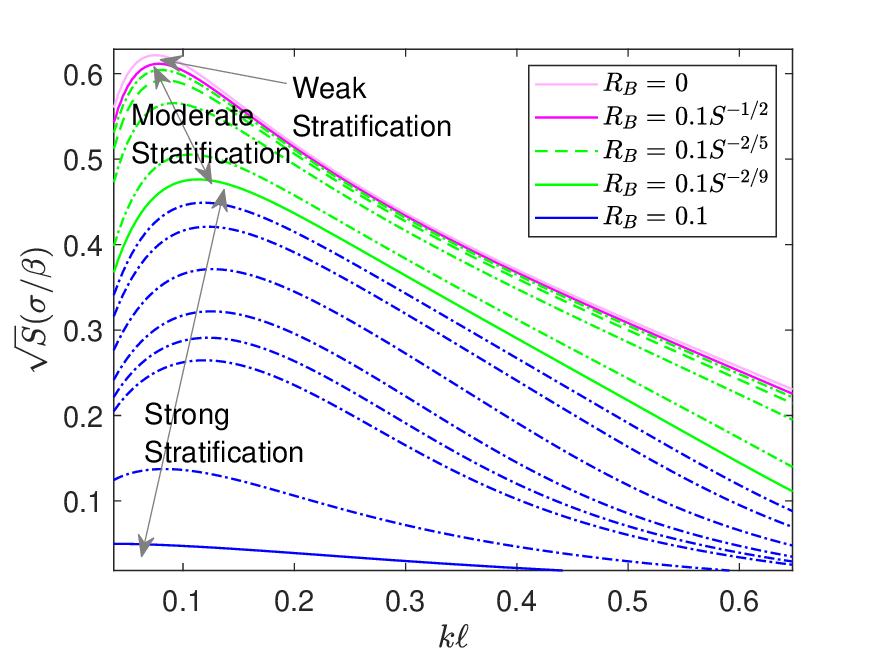}
    \caption{The dispersion relation~\eqref{eq:DispersionRelation_Full} for $S=10^5$ and various values of $R_B$.
    Solid, thick lines represent the transition between parameter regimes identified analytically:
    weak (magenta), moderate (green), strong (blue).
    The dot-dashed lines represent intermediate values of $R_B$ (not shown in legend).}
    \label{fig:FiniteSVariedRBDispRelPlots}
\end{figure}

\subsection{Finite Lundquist number, $S$} \label{subsec:Finite_S}

The analysis in the previous subsections considered the asymptotic limit $S\to\infty$,
but the same trends can be observed by plotting the dispersion relation~\eqref{eq:DispersionRelation_Full} for different values of $R_B$ with large but finite $S$.
Figure~\ref{fig:FiniteSVariedRBDispRelPlots} shows results for various values of $R_B$ in the case with $S = 10^5$.
As predicted analytically, we find that
as $R_B$ is increased
the peak shifts to larger values of $k$ once $R_B \gtrsim S^{-1/2}$,
and broadens and shifts back to smaller values of $k$ once $R_B \gtrsim S^{-2/9}$.

\section{Numerical validation}\label{sec:NumericalWork}

The results presented in the previous section are formally valid only in the asymptotic limit $S \to \infty$,
and also assume an unbounded domain.
In order to test the robustness of these analytical results,
in the presence of boundaries,
we have used a numerical eigensolver to obtain solutions of the linearised equations~\eqref{eq:FullODE_I} and \eqref{eq:FullODE_II} over a domain of finite size in $z$. The solver is based on the 
Newton--Raphson--Kantorovich (NRK) code originally developed by \citet{Gough76NRK}.
We take the numerical domain to be $z \in [0,100\ell]$
and impose that the perturbations $\hat{u}_{z}$ and $\hat{b}_{z}$ are antisymmetric and symmetric at $z=0$, respectively. At the other boundary, $z = 100\ell$, we impose that $\hat{u}_{z}$ and $\hat{b}_{z}$ both vanish. 

\begin{figure}
    \centering 
    \includegraphics[scale=0.6]{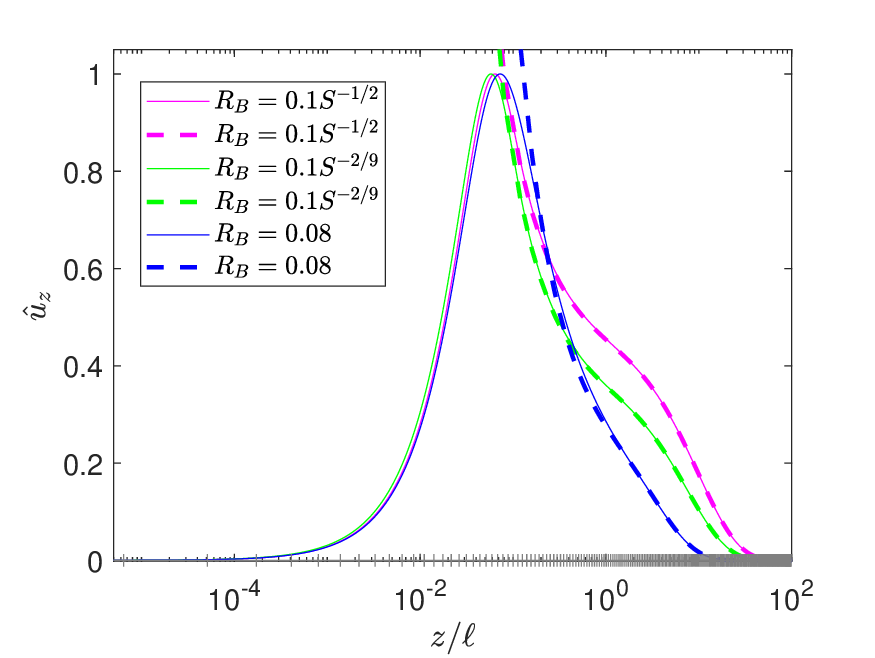}
    \caption{The numerical (solid lines) eigenfunction $\hat{u}_{z}(z)$, scaled to its peak value, for $k = 0.1/\ell$ and $S=10^5$, alongside the corresponding analytical (dashed lines) bulk $\hat{u}_{z}$ solution \eqref{eq:Bulk_Solution}, for various values of $R_B$.
    The horizontal axis is logarithmically scaled to give equal prominence to the boundary layer.
    The $+$ markers along this axis indicate each 20\textsuperscript{th} computational grid point out of 5000.
    The plots in the case $R_B = 0$ (not shown) are virtually indistinguishable from those for $R_B = 0.1S^{-1/2}$.}
    \label{fig:Eigenfunctions}
\end{figure}

In order to resolve both the bulk and boundary layer, we use a non-uniform grid with 5000 grid points spaced cubically in $z$,
i.e.~we have grid points at $z_n = 100\ell(n/5000)^3$.
Some example solutions for $\hat{u}_{z}$, for varying degrees of stratification,
are plotted in figure~\ref{fig:Eigenfunctions}.
We note that the main effect of increasing the stratification
is to suppress the perturbations in the bulk of the domain,
so that $\hat{u}_{z}$ becomes increasingly localised within the boundary layer.
Within the bulk, we find that $\hat{u}_{z}$ decays exponentially for large $|z|$, at the rate $\exp(-s|z|/\ell)$ predicted by equation~\eqref{eq:Bulk_Solution} (dashed lines in figure~\ref{fig:Eigenfunctions}). The exact choice of boundary conditions at $z = 100\ell$ should therefore have negligible effect for wavenumbers $|k| \gtrsim 1/(100\ell)$,
but may have an effect for smaller wavenumbers.
In the results that we present below we have taken the Lundquist number to be $S \leqslant 10^6$, anticipating that the fastest growing tearing mode will have $|k| > 1/(100\ell)$ and will therefore be insensitive to the boundary conditions. 
Conversely, for larger wavenumbers ($|k| \gtrsim 0.5/\ell$) the bulk solution decays rapidly away from the boundary layer,
becoming so small that rounding errors in the numerical solver become significant.
For this reason we limited the domain size to $100\ell$.

\subsection{Results} \label{subsec:NumericalResults}

\subsubsection{No stratification} \label{subsubsec:ClassicalComparison}
\begin{figure}
    \centering
    \includegraphics[scale=0.6]{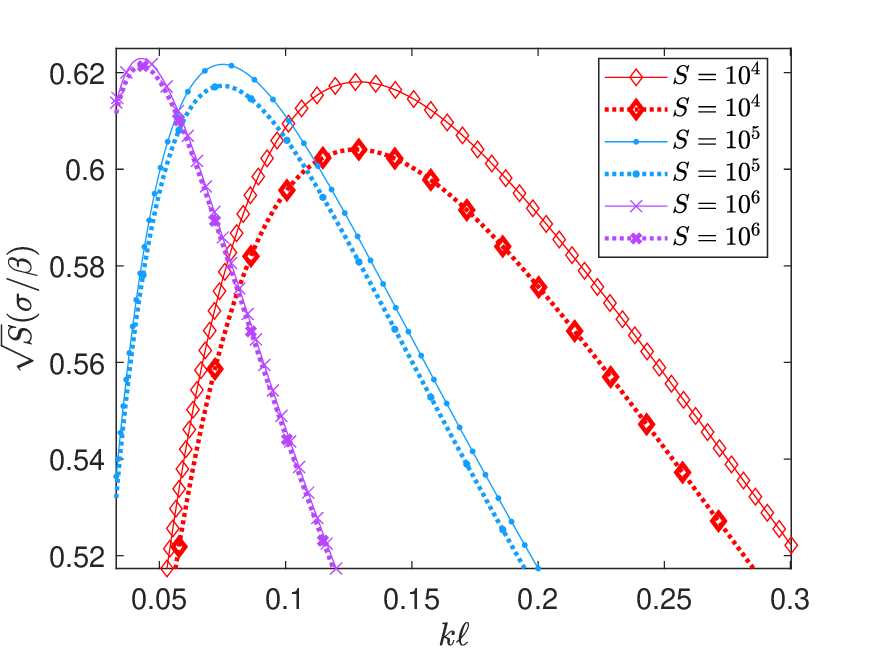}
    \caption{Analytical (solid lines) and numerical (dotted lines) dispersion relations for $R_B = 0$ (no stratification) for various values of $S$.
    }
    \label{fig:AnNumUnstratifiedPeaks}
\end{figure}

We first verify that the results from the numerical solver are consistent with our asymptotic solutions in the absence of stratification, i.e.~with $R_B = 0$.
Figure~\ref{fig:AnNumUnstratifiedPeaks} compares the analytical and numerical dispersion relation
for $S=10^4$, $10^5$ and $10^6$.
We find that the analytical result slightly overestimates the growth rate, but becomes increasingly accurate for larger $S$, as expected.
Even for $S=10^4$ however, the analytical result predicts the growth rate and the wavenumber of the fastest growing mode to within about $2\%$.

\subsubsection{Weak stratification} \label{subsubsec:a=1/2_Comparison}

We next consider the regime $S \to \infty$ with $R_B \sim S^{-1/2}$,
which represents the upper end of the ``weak stratification'' regime identified in section~\ref{subsec:Weak}.
\begin{figure}
    \centering
    \includegraphics[scale=0.6]{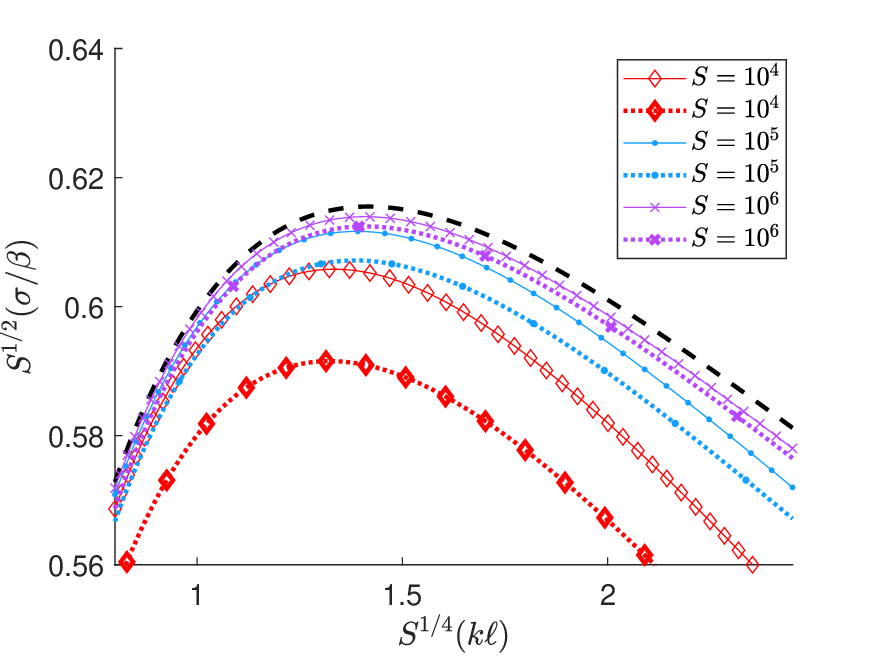}
    \caption{Analytical (solid lines) and numerical (dotted lines) dispersion relations for various values of $S$,
    with $R_B = 0.1S^{-1/2}$.
    The dashed curve shows the asymptotic solution~\eqref{eq:a1_2_DispRel} obtained in the limit $S\to\infty$.
    }
    \label{fig:AnNum_a=1/2}
\end{figure}
Figure~\ref{fig:AnNum_a=1/2} compares the analytical and numerical results for the dispersion relation for $S=10^4$, $10^5$ and $10^6$ with $R_B = 0.1S^{-1/2}$.
The axes in this figure are scaled with $S$, in order to match our analytical prediction for the fastest growing mode in the limit $S \to \infty$.
As in the unstratified case, the analytical result slightly overestimates the true growth rate, but becomes more accurate for larger $S$.
Moreover, for large $S$ the dispersion relation in a neighbourhood of the fastest growing mode converges to the result~\eqref{eq:a1_2_DispRel},
which is indicated by the dashed curve in figure~\ref{fig:AnNum_a=1/2}.

\subsubsection{Moderate stratification} \label{subsubsec:a=2/9_Comparison}
We next consider the regime $S \to \infty$ with $R_B \sim S^{-2/9}$,
which represents the upper end of the ``moderate stratification'' regime identified in section~\ref{subsec:Moderate}.
\begin{figure}
    \centering
    \includegraphics[scale=0.6]{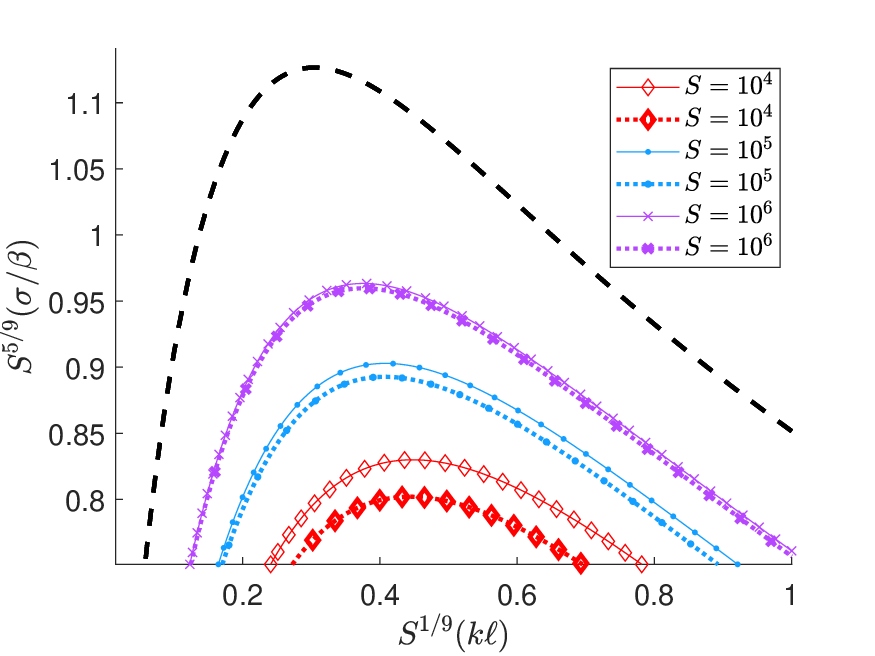}
    \caption{Analytical (solid lines) and numerical (dotted lines) dispersion relations for various values of $S$,
    with $R_B = 0.1S^{-2/9}$.
    The dashed curve shows the asymptotic solution~\eqref{eq:a2_9_DispRel} obtained in the limit $S\to\infty$.
    }
    \label{fig:AnNum_a=2/9}
\end{figure}
Figure~\ref{fig:AnNum_a=2/9} demonstrates the convergence of the analytical and numerical results for increasing values of $S$ in the case $R_B = 0.1S^{-2/9}$.
In this regime, the peak of the dispersion relation is well described by the formula~\eqref{eq:a2_9_DispRel}.

\subsubsection{Strong stratification} \label{subsubsec:a=0_Comparison}
For reasons discussed in section~\ref{subsec:Strong}, our analytical results cease to be valid when the stratification parameter $R_B$ is of order unity.
However, for any value of $R_B < 3/4$, our analytical result still offers a prediction for the wavenumber and growth rate of the fastest growing mode.
It is therefore interesting to compare this prediction with the dispersion relation obtained numerically.

\begin{figure}
    \centering
    \includegraphics[scale=0.6]{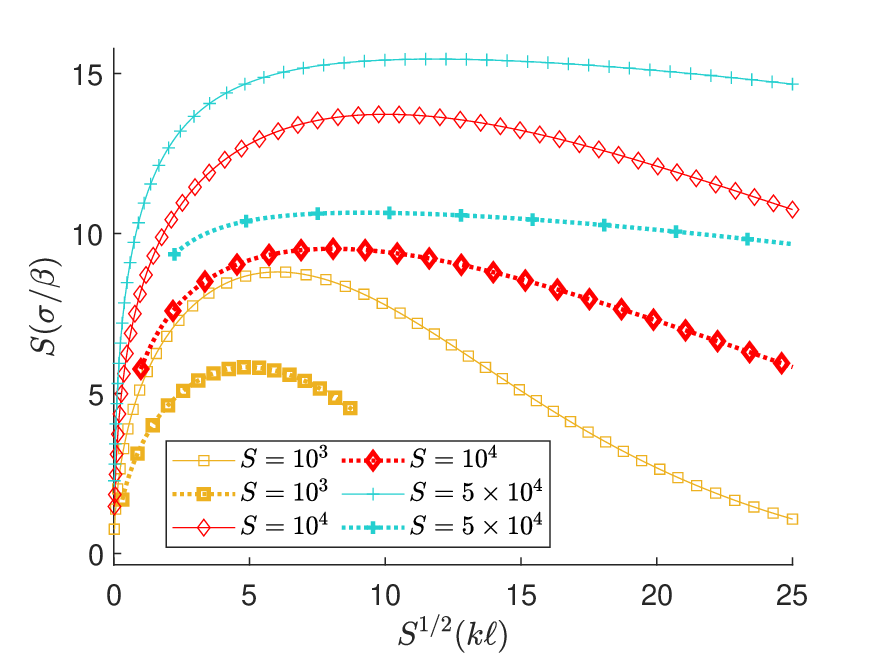}
    \caption{Analytical (solid lines) and numerical (dotted lines) dispersion relations for various values of $S$,
    with $R_B = 0.1$.
    }
    \label{fig:AnNum_a=0}
\end{figure}

As illustrated in figure \ref{fig:AnNum_a=0}, the numerical results show the same flattening of the dispersion curve predicted analytically, which becomes more pronounced for larger values of $S$.
The numerical results are also consistent with the prediction that the growth rate is of order $\sigma \sim \beta/S$ in this regime.
However, the analytical result overestimates the growth rate, and in contrast to the cases presented earlier, this discrepancy increases for larger values of $S$.
The analytical result is therefore not applicable in this regime.
{For larger (fixed) values of $R_B$ we would expect the discrepancy to be even larger, becoming infinite for $R_B = 3/4.$}

\section{Application to the solar tachocline}
\label{sec:Tachocline_Application}

The solar tachocline is believed to harbour a strong toroidal magnetic field,
amplified by rotational shear in that region.
The exact strength and topology of the field is highly uncertain,
but several studies have suggested values of order $10^4$\,G
\citep[e.g.][]{Antia00, Fan09, Jouve18}.
It has recently been argued that this toroidal field reverses in sign with depth, in the manner of a skin effect,
as a consequence of the cyclic nature of the solar dynamo
\citep{ForgacsDajkaPetrovay01,Barnabe-etal17,Matilsky-etal22}.
Such a configuration could potentially be subject to tearing instability,
and from our results we can attempt to estimate the growth rate of such instability.

We will adopt the same parameter values used by \citet{Barnabe-etal17}: a toroidal field of $B = 5\times10^{4}$\,G,
which reverses sign on a vertical scale of $\ell = 10^8$\,cm (about an order of magnitude smaller than the tachocline's thickness),
and a (turbulent) magnetic diffusivity of $\eta = 10^8$\,cm$^2$\,s$^{-1}$.
With these parameters (and taking the density to be $0.21$\,g\,cm$^{-3}$),
the Alfv\'enic shear is of order $\beta \sim 10^{-4}$\,s$^{-1}$,
which is somewhat smaller than the buoyancy frequency in the lower part of the tachocline, $N \sim 10^{-3}$\,s$^{-1}$,
implying a magnetic Richardson number of $R_{B} \sim 100$.
Therefore, based on our results, tearing instability can only operate in the upper, weakly stratified part of the tachocline. However, it must be admitted that our analysis has so far neglected the effect of thermal diffusion,
which would likely lessen the stabilising influence of stratification,
possibly allowing the instability to operate even in the deeper parts of the tachocline.
{Including thermal diffusion significantly complicates the problem from an analytical perspective,
but it is relatively straightforward to
include in our numerical eigensolver.
We therefore present a brief numerical analysis of the effects of thermal diffusion on instability below, in section~\ref{subsec:TDiff}.
First, however, we note that an upper bound for the tearing growth rate can be found by neglecting stratification entirely.
For the parameter values listed above,
the Lundquist number is $S \equiv \beta\ell^2/\eta \sim 10^4$,
implying an upper bound of
$\sigma \sim \beta S^{-1/2} \sim 10^{-6}$\,s$^{-1}$,
corresponding to growth times of the order of weeks.}
This is far shorter than the 22-year period of the solar dynamo cycle,
{so it seems unlikely that such a field could persist} in the upper, unstratified part of the tachocline.
Even in the absence of other disturbing influences, such as overshooting convective plumes,
we would expect an initially axisymmetric toroidal field to break up into magnetic islands via tearing instability,
much faster than it could diffuse down through the tachocline.

\subsection{Inclusion of thermal diffusion} \label{subsec:TDiff}

In the presence of thermal diffusion,
equation~(\ref{eq:Buoyancy_Strat}) becomes
\begin{equation}
  \frac{\partial\theta}{\partial t} + \boldsymbol{u} \boldsymbol{\cdot}\boldsymbol{\nabla}\theta = -N^{2}u_{z}
  + \kappa\nabla^2\theta,
\end{equation}
where $\kappa$ is the thermal diffusivity.
In the solar tachocline,
the \emph{microscopic} diffusivity of temperature
exceeds that of magnetic field
by several orders of magnitude
\citep[e.g.][]{Gough07}.
However, since we have employed the same turbulent value $\eta = 10^8$\,cm$^2$\,s$^{-1}$ used by \citet{Barnabe-etal17},
it seems appropriate to employ a similar value for $\kappa$
(which in any case is not much larger than the microscopic value).
It what follows we will therefore take
$\kappa = \eta$ for simplicity.

By the same process described in section~\ref{subsec:LinPerts}
we can reduce the linearised equations to a system of ordinary differential equations:
\begin{gather}
    \label{eq:TDiffODE_I}
     \sigma^2\hat{u}_{z}'' - \sigma^2k^2 \hat{u}_{z} + k^2\sigma\hat{\theta}
     = \frac{\mathrm{i} k\sigma}{4\pi\rho_0}\left[B\hat{b}_{z}'' - \left(k^{2} B + B''\right)\hat{b}_{z}\right],
     \\
     \label{eq:TDiffODE_II}
     \eta\hat{b}_{z}'' - \left[\eta k^2 + \sigma\right]\hat{b}_{z} = - \mathrm{i} kB\hat{u}_{z},
     \\
     \label{eq:TDiffODE_III}
     \kappa\hat{\theta}'' -\left[\sigma + k^2 \kappa\right]\hat{\theta} = N^2 \hat{u}_{z}.
\end{gather}

\begin{figure}
    \centering
    \includegraphics[scale = 0.6]{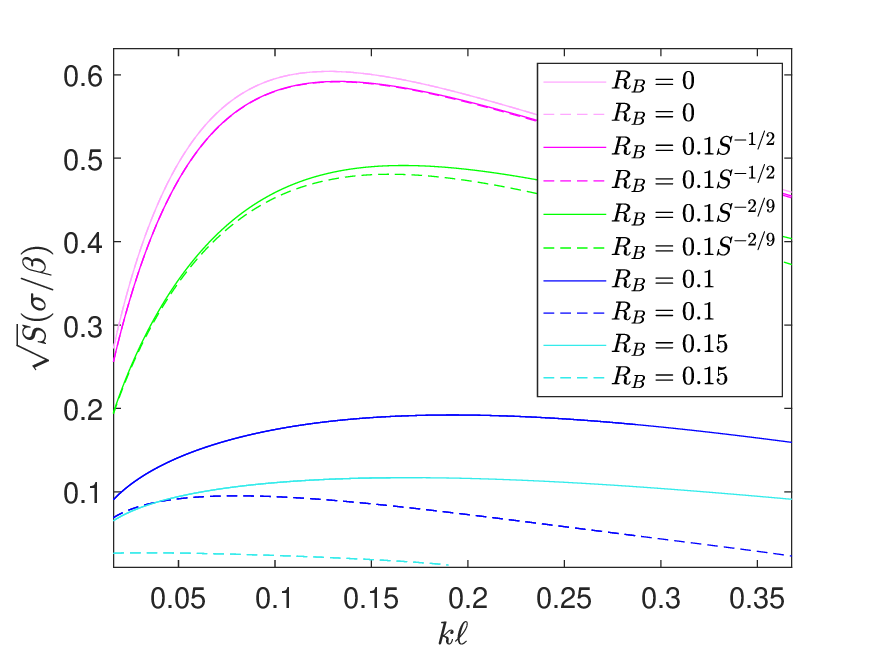}
    \caption{Numerical dispersion relations for $S=10^4$ and various values of $R_B$.
    Solid lines are used for $\kappa=\eta$
    and dashed lines are used for $\kappa=0$.
    In the unstratified case ($R_B=0$) the solid and dashed curves are identical.
    In the weakly stratified case ($R_B = 0.1S^{-1/2}$) the solid and dashed curves are virtually indistinguishable.}
    \label{fig:TDiffPlots}
\end{figure}

We have adapted the numerical eigensolver described in section~\ref{sec:NumericalWork}
to solve this system of equations,
with the additional boundary condition that $\hat{\theta}=0$ at $z=0$ and at the outer boundary.
Figure~\ref{fig:TDiffPlots}
compares the dispersion relations obtained numerically in the cases $\kappa=0$ and $\kappa=\eta$,
for $S=10^4$ and for various values of $R_B$.
(To reduce the computational burden
some of these results were obtained with a domain size of $80\ell$ and 1000 grid points.
We have verified that using a larger domain or additional grid points does not noticeably affect the results.)
In all cases with non-zero stratification,
including thermal diffusion leads to a larger growth rate,
and the fastest growing mode is found at smaller length scales (i.e.~larger $k$).
Physically, this reflects the fact that thermal diffusion acts to lessen the stabilising effect of the stratification,
especially on small scales.
This effect becomes more significant as the strength of the stratification (measured by $R_B$) is increased.
Interestingly, the results in figure~\ref{fig:TDiffPlots} suggest that,
with $\kappa=\eta$,
the fastest growing mode has $k\ell$ and $\sqrt{S}(\sigma/\beta)$ both of order unity even for ``strong'' stratification (i.e.~for $R_B$ of order unity).
Further work will be needed to confirm this, however,
and to determine whether tearing instability could occur in the deeper parts of the tachocline, where $R_B \sim 100$.

\section{Conclusion} \label{sec:Conclusion}

We have determined the effect of stable stratification on the resistive tearing instability.
{In the absence of thermal diffusion, the dispersion relation has been obtained both analytically and numerically.}
As the strength of the stratification is increased, it first suppresses perturbations in the bulk, so the instability becomes more localised to the boundary layer.
At the same time, the fastest growing mode shifts to smaller scales.
For stronger stratification, the smallest scales in the boundary layer are suppressed also,
and for sufficiently strong stratification the fastest growing mode shifts back to larger scales,
while the perturbations become even more localised within the boundary layer.
As the buoyancy frequency approaches the Alfv\'enic shear rate, the growth rate drops to that of bulk magnetic diffusion,
and the instability is effectively nullified.

Our dispersion relation~\eqref{eq:DispersionRelation_Full} generalises the well-known unstratified result~\eqref{eq:UnstratifiedDispRel} to all values of stratification in the range $0 \leqslant R_B \ll 1$,
and the right-hand side reduces to the result of \citet{Johnson63} in what we have called the ``moderately stratified regime", wherein stratification begins to affect the boundary layer.
Because our dispersion relation ceases to be valid for $R_B \gtrsim 1$ (at which point the tearing instability is effectively nullified),
it can be simplified somewhat by assuming that $R_B \ll 1$.
The simplest form that is valid for all values of $k$ is
\begin{equation}
  (Sk\ell)^{1/3}\frac{\sqrt{(k\ell)^2+R_B}}{1-(k\ell)^2}
  =
  \frac{1-\lambda^2}{\pi\lambda^{5/6}}\frac{\Gamma\left(\frac{(1+\lambda)(R_B+\lambda)}{4\lambda}\right)}{\Gamma\left(\frac{(3+\lambda)(\tfrac{1}{3}R_B+\lambda)}{4\lambda}\right)}\,.
\end{equation}
It is straightforward to check that this simplified form is consistent with all of the results~\eqref{eq:UnstratifiedDispRel}--\eqref{eq:a2_9_DispRel} presented in section~\ref{sec:AsympMatching}.

{In the presence of thermal diffusion, we have obtained the dispersion relation numerically.
We find that thermal diffusion generally leads to a faster growing instability that operates on smaller length scales,
and that this effect becomes increasingly significant as the strength of the stratification increases.
On the basis of these results, we conclude that the alternating, axisymmetric toroidal field in the tachocline proposed in some models \citep{ForgacsDajkaPetrovay01,Barnabe-etal17}
would likely be subject to non-axisymmetric tearing instability on a timescale far shorter than the solar cycle.}
This conclusion is supported by recent numerical simulations \citep{Matilsky-etal24},
which suggest that the field in the tachocline is characterised by long-lived, non-axisymmetric structures,
rather than a cyclic, axisymmetric toroidal field.
\\

TSW and PJB were supported by a Research Project Grant from the Leverhulme Trust (RPG-2020-109).
TSW also acknowledges funding from the Science and Technology Facilities Council (ST/W001020/1).
\\

\noindent \textbf{Declaration of interests:} The authors report no conflict of interests.
\\

\noindent \textbf{Author ORCIDs:} S. J. Hopper, https://orcid.org/0009-0004-3835-3888; T. S. Wood, https://orcid.org/0000-0003-1044-170X; P. J. Bushby, https://orcid.org/0000-0002-4691-6757

\appendix

\section{Frobenius series} \label{sec:Frobenius}
In order to asymptotically match the bulk solution~\eqref{eq:Bulk_Solution}
to the boundary-layer solution~\eqref{eq:BL_Solution},
we must first express them both in the form of Frobenius series,
as in equations~\eqref{eq:Bulk_Frobenius} and \eqref{eq:BL_Frobenius}.

The behaviour of the bulk solution~\eqref{eq:Bulk_Solution} for small $z$ can be deduced using the identity
\begin{align}
  {}_2F_1(a,b;c;z) = {} & \frac{\Gamma(c)\Gamma(c-a-b)}{\Gamma(c-a)\Gamma(c-b)}{}_2F_1(a,b;a+b+1-c;1-z) \nonumber \\
  & {} + \frac{\Gamma(c)\Gamma(a+b-c)}{\Gamma(a)\Gamma(b)}(1-z)^{c-a-b} {}_2F_1(c-a,c-b;1+c-a-b;1-z),
\end{align}
which allows us to express the solution as a power series in $T$, and hence $z$.
We thus eventually arrive at a power series in the form of equation~\eqref{eq:Bulk_Frobenius}, with
\begin{align}
  A_0 &= \frac{\Gamma(r)\Gamma(1+s)\ell^{\tfrac{1}{2}+r}}{\Gamma(\tfrac{5}{4}+\tfrac{1}{2}s+\tfrac{1}{2}r)\Gamma(-\tfrac{1}{4}+\tfrac{1}{2}s+\tfrac{1}{2}r)}\,,
  \nonumber \\
  B_0 &= \frac{\Gamma(-r)\Gamma(1+s)\ell^{\tfrac{1}{2}-r}}{\Gamma(\tfrac{5}{4}+\tfrac{1}{2}s-\tfrac{1}{2}r)\Gamma(-\tfrac{1}{4}+\tfrac{1}{2}s-\tfrac{1}{2}r)}\,.
  \label{eq:Bulk_coefficients}
\end{align}
We note that, if $r$ is an integer, then at least one of these coefficients is undefined.
This is because the Frobenius series then involves logarithmic terms,
as described in section~\ref{sec:Bulk}.

The boundary-layer solution~\eqref{eq:BL_Solution},
which is defined in Fourier space,
can readily be expressed as a power series in $X$, and hence $\zeta$.
This solution has the form
\begin{equation}
  \uztilde = \sum_{n=0}^{\infty}\left[\tilde{a}_{n}|\zeta|^{-\frac{1}{2}+r+2n} + \tilde{b}_{n}|\zeta|^{-\frac{1}{2}-r+2n}\right]\sgn(\zeta),
  \label{eq:BL_Fourier}
\end{equation}
where
\begin{gather}
  \tilde{a}_0 = \frac{2\lambda\Gamma(-r)}{\left(\lambda-r+\tfrac{1}{2}\right)\Gamma\left(\dfrac{(\lambda-r)^2-\tfrac{1}{4}}{4\lambda}\right)}\left(\dfrac{\sqrt{\lambda}\ell^2}{Sk}\right)^{-\tfrac{1}{6}+\tfrac{1}{3}r}\,,
  \nonumber \\
  \tilde{b}_0 = \frac{2\lambda\Gamma(r)}{\left(\lambda +r + \tfrac{1}{2}\right)\Gamma\left(\dfrac{(\lambda+r)^2-\tfrac{1}{4}}{4\lambda}\right)}\left(\dfrac{\sqrt{\lambda}\ell^2}{Sk}\right)^{-\tfrac{1}{6}-\tfrac{1}{3}r}\,.
\end{gather}
We now take the inverse Fourier transform,
using the identity~\eqref{eq:FT_IFT_uz}.
We thus arrive at a power series in the form of equation~\eqref{eq:BL_Frobenius}, with
\begin{equation}
  a_0 = \frac{\tilde{a}_0}{2\mathrm{i}\Gamma(\tfrac{1}{2}-r)\sin[\tfrac{\pi}{2}(\tfrac{1}{2}-r)]}
  \quad \mbox{and} \quad
  b_0 = \frac{\tilde{b}_0}{2\mathrm{i}\Gamma(\tfrac{1}{2}+r)\sin[\tfrac{\pi}{2}(\tfrac{1}{2}+r)]}\,.
\end{equation}
Substituting the above expressions for the coefficients $A_0$, $B_0$, $a_0$ and $b_0$ into equation~\eqref{eq:CoeffDispRel}
yields the dispersion relation given in equation \eqref{eq:DispersionRelation_Full}.

{We have assumed throughout that $\hat{u}_{z}(z)$
is an odd function, and hence so is $\uztilde(\zeta)$.
For completeness, we will now consider the opposite case,
in which $\hat{u}_{z}(z)$ and $\uztilde(\zeta)$ are both even functions.
In that case, the solution proceeds just as before,
except for the absence of the factors $\sgn(z)$ and $\sgn(\zeta)$
in equations~\eqref{eq:Bulk_Solution}, \eqref{eq:BL_Solution}, \eqref{eq:Bulk_Frobenius}, \eqref{eq:BL_Frobenius} and \eqref{eq:BL_Fourier}.
However, when we take the inverse Fourier transform of equation~\eqref{eq:BL_Fourier}, in place of the identity~\eqref{eq:FT_IFT_uz}
we must now use \citep[see][]{Kammler08}
\begin{equation}
  \uztilde(\zeta) \sim |\zeta|^{-\alpha} \quad \mbox{as $\zeta\to0$}
  \qquad \Longleftrightarrow \qquad
  \hat{u}_{z}(z) \sim \frac{|z|^{\alpha-1}}{2\cos(\tfrac{\pi}{2}\alpha)\Gamma(\alpha)} \quad \mbox{as $|z|\to\infty$}
\end{equation}
which ultimately leads to the same dispersion relation~\eqref{eq:DispersionRelation_Full}, except with both sine functions replaced by cosines.
For any value of $r$ in the interval $\tfrac{1}{2} < r \leqslant 1$
this change introduces a small numerical factor to the right-hand side,
which is equivalent to increasing the value of $S$, and results in a smaller growth rate.
In the regime $R_B \ll 1$ this small factor on the right-hand side is approximately $(\pi R_B/2)^2$,
and we find that the growth rate now predicted by the dispersion relation is much smaller than the rate of bulk diffusion, i.e.~$\sigma \ll \beta/S$.
Therefore we conclude that tearing instability is effectively absent for perturbations that have an even $\hat{u}_{z}(z)$.}

\bibliographystyle{jfm}
\bibliography{Bibliography}

\end{document}